\documentclass[letterpaper, 10 pt, conference]{ieeeconf}  
\usepackage{amsfonts}
\usepackage{amssymb}
\usepackage{amsmath}                                                        
\usepackage{subfig}
\usepackage{epsfig}
\usepackage{graphicx}
\IEEEoverridecommandlockouts                              
\newtheorem{theorem}{Theorem}
\newcommand{\startproof}{\vspace{0.2cm} \noindent {\bf \em Proof: }}   
\newcommand{\finishproof}{\hfill $\blacksquare$}    
\newtheorem{assumption}{Assumption}                                                   
   
\newtheorem{remark}{Remark}                                                      
\title{\LARGE \bf
A Minimax Linear Quadratic Gaussian Method for Antiwindup Control Synthesis}

\author{Obaid ur Rehman, Ian R. Petersen and Bar\i\c{s} Fidan
\thanks{Obaid Ur Rehman is with school of Engineering and Information Technology, University of New South Wales at the Australian Defence Force Academy, Canberra, Australia. ({\tt\small s.obaid.rehman@gmail.com}}%
\thanks{Prof. Ian R. Petersen is with school of Engineering and Information Technology, University of New South Wales at the Australian Defence Force Academy, Canberra, Australia. ({\tt\small i.r.petersen@gmail.com})}%
\thanks{Bar\i\c{s} Fidan is with department of Mechanical and Mechatronics Engineering, University of Waterloo, Canada ({\tt\small fidan@uwaterloo.ca})}
}

\begin{document}

\maketitle
\thispagestyle{empty}
\pagestyle{empty}

\begin{abstract}

In this paper, a dynamic antiwindup compensator design is proposed which augments the main controller and guarantees robust performance in the event of input saturation. This is a two stage process in which first a robust optimal controller is designed for an uncertain linear system which guarantees the internal stability of the closed loop system and provides robust performance in the absence of input saturation. Then a minimax linear quadratic Gaussian (LQG) compensator is designed to guarantee the performance in certain domain of attraction, in the presence of input saturation. This antiwindup augmentation only comes into action when plant is subject to input saturation. In order to illustrate the effectiveness of this approach, the proposed method is applied to a tracking control problem for an air-breathing hypersonic flight vehicle (AHFV).
\end{abstract}

\section{INTRODUCTION}

The design of a controller for a linear system with input saturations is a challenging task. Input saturation can have a disastrous effects \cite{AW_Doyle}. In general, it is common to consider control problem involving integral action and in the presence of input saturations, the state of the integrator can wind up and affect the response of the underlying system. Engineers using PI and PID control have solved the windup problems using ad hoc measures, such as by resetting the integral states in the case of windup, or by using a modified digital implementation of the controller, which does not depend on the integral error terms. However, these ad hoc measures lack mathematical rigor and are mostly heuristic \cite{AW_Astrom, AW_Hanus}. Attempts have also been made by control system designers to penalize the control output so that actuator limits would never be violated. These method are useful to some extent but may result in a design which is too conservative and unable to utilize the full capability of the available control authority. 

In order to use full controller authority of the controller and to recover the nominal performance of the system in the event of actuator saturation, considerable attention has been given to antiwindup augmentation over the last decades. The main advantage of using an antiwindup augmentation scheme is that the system will operate with its full capability in term of robustness and performance in the absence of input saturation. The antiwindup correction only come into effect in an event of input saturation. Antiwindup augmentation, as the name suggests, is a two-stage design procedure. In this design procedure, the requirement of small signal behavior of the nominal system is guaranteed by ignoring the saturation. Then antiwindup compensation is added to recover the nominal performance in the presence of saturation. 

Noting the importance of the antiwindup augmentation to handle saturation, different methods have been appeared in the literature over the decades. Most of the methods relying on the $\mathcal{H}_{\infty}$ optimal control approach have considered the antiwindup problem as an $\mathcal{L}_2$ gain minimization problem \cite{AW_Edwards, AW_Unstable_WU, AW_Convex_Ferreres, AW_LMI_Mulder, AWP01}. In these papers, saturation nonlinearities are considered as a sector or dead-zone type nonlinearities and the antiwindup problem is solved by formulating it in a convex optimal framework. In \cite{AW_Convex_Ferreres, AW_LPV_Wu, AW_LFT_Wu}, linear fractional transformation (LFT) and linear parameter varying (LPV) approaches have been proposed for parameter-varying saturated systems. All of these methods employs linear matrix inequalities (LMIs) as a tool to compute global optima in a simplified way. The main drawback of the LMI framework is that in several situations LMI constraints are unfeasible and hence no solution exists. Also, LMI methods can suffer from numerical and computational problems in the case of high order system.  These problems are solved to some extent in \cite{AWP01} by characterizing the antiwindup problem in terms of a nonconvex feasibility problem, which reduces to a convex feasibility problem when a certain rank constraint becomes inactive. Another solution of this problem is attempted in \cite{AW_QLF_Mulder} where, the non-feasible problem was solved based on the approximate solution of an LMI. The results of \cite{AWP01,AW_QLF_Mulder} are only applicable to linear stable plants. Antiwindup controller design for an unstable system was considered in \cite{AW_Unstable_WU} which is also based on the LMI method. However, the results are only valid when no uncertainty exists in the original system except for the uncertainty due to the actuator saturation.

In this paper we have considered stabilizable unstable linear systems subject to parameter uncertainties and input saturations. The formulation of the antiwindup problem is based on an $\mathcal{H}_{\infty}$ framework. In this framework a dynamic antiwindup compensator is proposed, which solves a  risk-sensitive control problem using the minimax LQG (a robust version of LQG control) design method \cite{IP}. This is similar to the $\mathcal{L}_2$ gain optimization problem. The only difference is that in minimax approach, we minimize the upper bound on a time averaged cost function and guarantee the robust performance. This method not only allows for the uncertainty which arises from saturation nonlinearities but also accounts for uncertainties present in the original system. In the propose method, a suitable robust controller is obtained by ignoring actuator saturation and then the closed loop system is augmented by a robust antiwindup compensator. 

The proposed antiwindup controller is then applied to solve the actuator saturation problem in an air-breathing hypersonic flight vehicle. The control problem in this example poses significant challenges as it offers a challenging trade-off among conflicting requirements. In high speed aircraft, the model is subject to parameter uncertainties due to a large flight envelop and a good robust controller is required to be designed. However, in the absence of antiwindup augmentation, limited control authority seldom allows the robust controller to work in its full capacity and thus degradation in the performance is unavoidable. The antiwindup compensator proposed in this paper solves these problems quite effectively and can be easily implemented on-board the aircraft.

The paper is organized as follows. Section \ref{sec:system} describes the class of uncertain linear systems and uncertainties considered in the paper. Section \ref{sec:awproblem} describes the general antiwindup problem and the synthesis of an antiwindup compensator using the minimax LQG method. The application of the proposed method to an AHFV control problem along with simulation results are presented in Section \ref{sec:example}. The paper is concluded in Section \ref{sec:concl}, with some final remarks on the proposed procedure.

\section{System Definition}\label{sec:system}
Consider an unconstrained uncertain linear unstable plant given by

\begin{equation}
\label{eqgform1}
\begin{split}
\dot {x}(t) &=A x(t)+B u(t)+\sum^l_{j=1} C_j\zeta_ j;\\
z_i(t)&=K_i{x(t)}+G_i u(t);\quad i=1,2,\cdots,m\\
y(t)&=\bar{C}_2 x(t)+ D_2 u(t)
\end{split}
\end{equation}
where $x(t)\in \mathbb{R}^n$ is the plant state, $u(t)\in\mathbb{R}^{n_u}$ is the control input, $\zeta_ j\in \mathbb{R}^{n_{\zeta}}$ is the uncertainty input, $z_i(t)\in \mathbb{R}^{n_{q}}$ is the uncertainty output, $y\in \mathbb{R}^{n_{y}}$ and $A$, $B$, $C_j$, $K_i$, $G_i$, $\bar{C}_2$, and $D_2$ are the matrices of suitable dimensions. Assume also that the uncertainty in the system satisfy following integral quadratic constraint condition (IQC) \cite{IP} 
\begin{equation}
\label{eqIQC}
\int_0^{\infty}(\parallel z_j(t)\parallel^2-\parallel \zeta_j(t)\parallel^2) dt \geq-x^T(0) d_j x(0),
\end{equation}
where $d_j>0$ for each $j=1,\cdots,l$ is a given positive definite matrix. Also, assume a minimax optimal linear quadratic regulator (LQR) control \cite{MMX01} of the following form exists for the system (\ref{eqgform1}) which is well posed and guarantees internal stability of the closed loop system:
\begin{equation}
\label{eqcontrollaw}
u(t)=-G_\tau x(t),
\end{equation}
where
\[
G_\tau=E^{-1}[B^T X_\tau+G^T K].
\]

Here, $X_\tau$ is obtained by solving a game type Riccati equation 
\begin{align}
\label{eqARE}
&(A-BE^{-1}G^T K)^T X_\tau+X_\tau(A-BE^{-1}G^T K)\nonumber\\
&+X_\tau(CC^T-BE^{-1}B^T)X_\tau+K^T(I-GE^{-1}G^T)K=0,
\end{align}
where
\[
K=\left[\begin{array}{c}
Q^{1/2} \\
0 \\
\sqrt{\tau_1}K_1 \\
\vdots\\
\sqrt{\tau_l}K_l
\end{array}\right]
,\quad
G=\left[\begin{array}{c}
0 \\
R^{1/2} \\
\sqrt{\tau_1}G_1 \\
\vdots\\
\sqrt{\tau_l}G_l
\end{array}\right]
,\quad
E=GG^T,
\]
\[
C=\left[\begin{array}{ccc}
\frac{1}{\sqrt{\tau_1}}C_1 & \dots & \frac{1}{\sqrt{\tau_l}}C_l
\end{array}\right]
\]
for given parameters $\tau_1>0,\tau_2>0$,...,$\tau_l >0$. The parameters $\tau_j$ for $j=1,\cdots,l$ are selected such that they give a minimum value of a bound on the following cost function 
\begin{equation}
\label{eqfcostlqr}
F=\int_0^{\infty}{[x(t)^T Q xt)+v(t)^T R v(t)]dt},
\end{equation}
where, $Q=Q^T>0$ and $R=R^T>0$ are the state and control weighting matrices respectively and the solution $X_\tau$ of the Riccati equation (\ref{eqARE}) should be symmetric and positive definite.  The bound on the cost function is given as
\begin{equation}
\label{eqfbond}
\text{min}[x^T(0)X_\tau x(0)+\sum_{j=1}^{l}\tau_j x^T(0)d_jx(0)],
\end{equation}
where the initial condition $x(0)\neq0\in \mathbb{R}^n $ is assumed to be known. Alternatively, the initial condition can be assumed to be a zero mean unity variance random vector, in which case the trace of the matrix in (\ref{eqfbond}) would be minimized. 
\begin{theorem}
\label{th1}
Consider the uncertain linear system (\ref{eqgform1}) with cost function (\ref{eqfcostlqr}). Then for any $\tau_1>0,\tau_2>0$,...,$\tau_l >0$ such that Riccati equation (\ref{eqARE}) has a positive definite solution, the  controller (\ref{eqcontrollaw}) is a guaranteed cost controller for this uncertain system with any initial condition $x(0) \in \mathbb{R}^{n}$. Furthermore the corresponding value of the cost function (\ref{eqfcost}) is bounded by the quantity (\ref{eqfbond}) for all admissible uncertainties and moreover, the closed loop system is absolutely stable.
\end{theorem}
\startproof
See \cite{IP,MMX01}.
\finishproof

\section{The Antiwindup Problem}\label{sec:awproblem}
\begin{figure}[t]
\hfill
\begin{center}
\epsfig{file=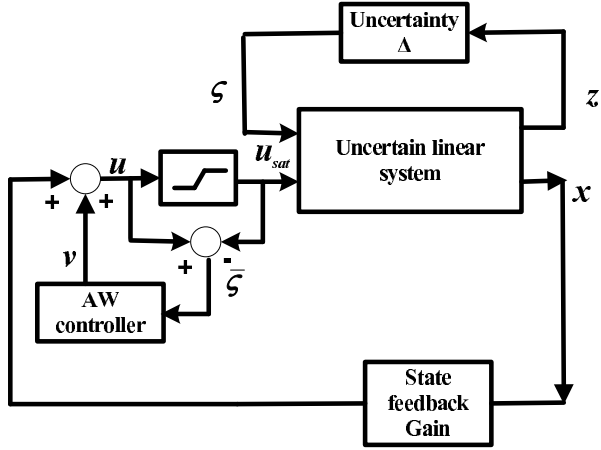, scale=0.8}
\caption{Closed loop system structure with antiwindup augmentation. }
\label{fig:awsystem}
\end{center}
\end{figure}

In real control problem, the input $u(t)$ is subject to saturation and thus the control law (\ref{eqcontrollaw}) may not give satisfactory performance. Indeed, each component $u_i$ of the input vector $u$ is subject to a saturation nonlinearity of the form shown in Fig. \ref{fig:Usat}. The antiwindup problem here is to design a suitable antiwindup compensator which guarantees adequate performance of the closed loop system in the presence of saturation nonlinearity. 
We represent saturation nonlinearity as a deadzone type sector bounded uncertainty and defines corresponding deadzone function $\phi_i(u_i)$ as follows:
\begin{equation}
\begin{split}
\phi_i(u_i)&=u_i-u_{i_{sat}}\\
&=\Bigg\{\begin{array}{c}
0, \quad \vert u_i \vert <  \vert u_{i_{max}} \vert  \\
sgn(u_i)(u_i-\vert u_{i_{max}}\vert), \quad \vert u_i \vert > \vert u_{i_{max}}\vert
\end{array}
\end{split}
\end{equation}
where, $u_{sat}$ is the input vector which is subject to saturation.
Also, we augment the controller in (\ref{eqcontrollaw}) with an antiwindup augmentation (see Fig. \ref{fig:awsystem}) as follows:
\begin{equation}
\label{eqcontrollawnew}
u(t)=-G_\tau x(t)+ v,
\end{equation}
where $v$ is the signal from the antiwindup compensator of the form
\begin{equation}
\begin{split}
\label{eqawcontrol}
\dot{x}_{aw}&=A_{aw} x_{aw}+B_{aw} \bar{\zeta},\\
v&=C_{aw} x_{aw},
\end{split}
\end{equation}
where $A_{aw}$, $B_{aw}$ and $C_{aw}$ are the matrices of suitable dimension and $\bar{\zeta}$ is the input to the compensator. The procedure to obtain these matrices will be given in sequel. 

\subsection{Antiwindup controller design} 
\begin{figure}[t]
\hfill
\begin{center}
\epsfig{file=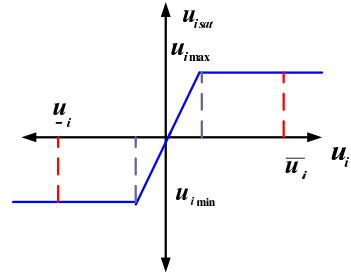, scale=0.4}
\caption{Control input saturation.}
\label{fig:Usat}
\end{center}
\end{figure}

\begin{figure}[t]
\hfill
\begin{center}
\epsfig{file=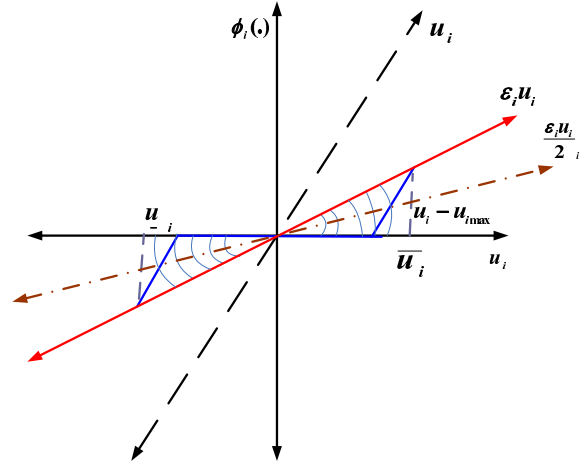, scale=0.4}
\caption{Deadzone uncertainty representation with domain of attraction.}
\label{fig:uregion}
\end{center}
\end{figure}

The design of an antiwindup augmentation system for the uncertain linear system is designed by defining a new equivalent uncertain model, considering the saturation nonlinearities as sector bounded uncertainties (deadzone type) and then froming an equivalent closed loop system using (\ref{eqcontrollawnew}). Firstly, we define a domain of attraction $\mathfrak{D_c}$ by restricting $u_i \in [\underbar{u}_i, \bar{u}_i]$ for $i=1,\cdots,m$ where, $\bar{u}_i$ is the maximum allowed value of $u_i$ and $\underbar{u}_i$ is the minimum allowed value of $u_i$. The corresponding  sector bound can be selected by appropriately choosing $0 < \epsilon_i \leq 1 $ (see Fig. \ref{fig:Usat} and Fig. \ref{fig:uregion})
and by using the following equation.
\begin{equation}
\begin{split}
\bar{u}_i-u_{i_{max}}&=\epsilon_i \bar{u}_i,\\
\bar{u}_i&=\frac {u_{i_{max}}} {1-\epsilon_i}
\end{split}
\end{equation}
Also, note that $u_{i_{min}}=-u_{i_{max}}$. 
The open loop system can be written using (\ref{eqgform1}) considering all input saturations as follows:
\begin{equation}
\label{eqgformnew}
\begin{split}
\dot{x}(t) &=A x(t)+B u_{sat}+\sum^l_{j=1} C_j\zeta_ j;\\
&=A x(t)+B (u-\phi(.))+\sum^l_{j=1} C_j\zeta_ j;\\
&=A x(t)+B u-B\phi(.)+\sum^l_{j=1} C_j\zeta_ j;\\
\end{split}
\end{equation}
where, $\phi(\cdot)=[\phi_i(u_i),\cdots, \phi_m(u_m)]^T$.

\begin{figure}[t]
\hfill
\begin{center}
\epsfig{file=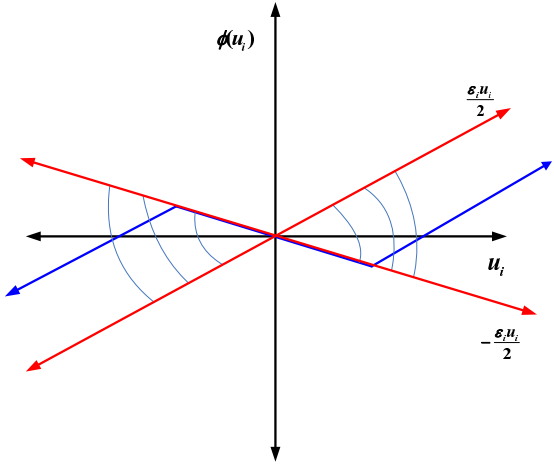, scale=0.4}
\caption{Uncertainty representation with new sector.}
\label{fig:uregion_norm}
\end{center}
\end{figure}

The sector bound on each $\phi_i(u_i)$ for $i=1,\cdots,m$ can be written as follows:
\begin{equation}
\begin{split}
0 &\leq \frac{\phi_i(u_i)}{u}\leq \epsilon_i\\
0 &\leq \phi(u_i) u_i \leq \epsilon_i u_i^2.
\end{split}
\end{equation}
We then define a new uncertainty input as
\begin{equation}
\begin{split}
\hat{w}_i(u_i)&=\phi_i(u_i)-\frac{\epsilon_i u_i}{2}\\
\Rightarrow~\hat{w}(u_i)&+\frac{\epsilon_i u_i}{2}=\phi_i(u_i),\\
\end{split}
\end{equation}
and write the uncertainty corresponding to this uncertainty input in a new sector as follows:
\begin{equation}
\label{eqruncertainty}
\begin{split}
0 \leq \hat{w}_i(u_i)u_i+\frac{\epsilon_i u_i^2}{2}&=\phi(u_i) u_i \leq \epsilon_i u_i^2\\
\Rightarrow -\frac{\epsilon_i u_i^2}{2}&\leq \hat{w}(u_i)u_i \leq \frac{\epsilon_i u_i^2}{2}.
\end{split}
\end{equation}
Also, the sector bound (\ref{eqruncertainty}) on the saturation uncertainty can be written in the following form (See Fig. \ref{fig:uregion_norm}). 
\begin{equation}
\label{eqnormbounded}
\Vert \hat{w}_i \Vert\leq \frac{\epsilon_i \vert u_i\vert}{2} =\vert \bar{z} \vert,
\end{equation}
where, $\bar{z}=\frac{\epsilon_i u_i }{2}$.

\begin{remark}
The formulation presented here  is applicable to the case where $\vert u_{i_{min}} \vert = \vert u_{i_{max}} \vert$; i.e. the symmetric saturation case (see Fig. \ref{fig:Usat}). However, it is straightforward to extend this formulation to the asymmetric saturation case where $\vert u_{i_{min}} \vert \neq \vert u_{i_{max}} \vert$.
\end{remark}

We can satisfy the bound (\ref{eqgformnew}) using the new uncertainty in (\ref{eqnormbounded}) as follows:
\begin{equation}
\label{eqformnormbound}
\begin{split}
\dot{x}(t)&=A x(t)+B u+\sum^l_{j=1} C_j\zeta_ j- B(\hat{w}(\cdot)+\frac{\mathfrak{E} u}{2})\\
\dot{x}(t)&=A x(t)+B(1-\frac{\mathfrak{E}}{2}) u+\sum^l_{j=1} C_j\zeta_ j- B \hat{w}
\end{split}
\end{equation}
where, $\hat{w}(\cdot)=[\hat{w}_1(u_1), \cdots, \hat{w}_m(u_m)]^T$, $\mathfrak{E}=diag[\epsilon_i, \cdots, \epsilon_m]$, $\bar{B}=B(1-\frac{\mathfrak{E}}{2})$, $\bar{G}$ is a scaling matrix corresponding to the bound in (\ref{eqnormbounded}) as given below:
\begin{equation}
\bar{G}=\left[\begin{array}{ccc}
\epsilon_i/2 & \cdots & 0 \\
0 & \ldots & 0 \\
0 & 0 & \epsilon_m/2
\end{array}\right].
\end{equation}

Since, we have considered additional uncertainties corresponding to the saturation uncertainties in (\ref{eqformnormbound}), the uncertainty outputs of the original system (\ref{eqgform1}) along with the uncertainty output $\bar{z}$ corresponding to $\hat{w}$ can be written as follows:
\begin{equation}
\label{eqgformznew}
\tilde{z}=\tilde{K}x+\tilde{G}u,
\end{equation}
where
\[
\tilde{z}=\left[\begin{array}{c}
z_1\\
z_2\\
\vdots\\
z_l\\
\bar{z}
\end{array}\right]; ~
 \tilde{K}=\left[\begin{array}{c}
K_1\\
K_2\\
\vdots\\
K_l\\
\mathbf{0}
\end{array}\right]; ~
\tilde{G}=\left[\begin{array}{c}
G_1\\
G_2\\
\vdots\\
G_l\\
\bar{G}
\end{array}\right]
\].

Also, we can write the complete uncertain linear model of the system in the presence of actuator saturation as given below:
\begin{equation}
\label{eqfinalform}
\begin{split}
\dot{x}(t)&=A x(t)+\bar{B}u+\sum^l_{j=1} C_j\zeta_ j- B \hat{w}(\cdot),\\
\tilde{z}&=\tilde{K}x+\tilde{G}u.
\end{split}
\end{equation}

The minimax optimal controller in (\ref{eqcontrollawnew}) can be used to obtain closed loop system which guarantees stability for the uncertain system (\ref{eqgform1}) without saturation uncertainty. The closed loop system incorporating the antiwindup signal $v$ in (\ref{eqfinalform}) can be written as follows:
\begin{equation}
\begin{split}
\dot{x}(t)&=A x(t)+\bar{B}(G x+v)+\sum^l_{j=1} C_j\zeta_ j- B \hat{w}(\cdot),\\
&=(A-\bar{B}G) x(t)+ \bar{B} v+\sum^l_{j=1} C_j\zeta_ j- B \hat{w}(\cdot).
\end{split}
\end{equation}
In a similar way using (\ref{eqcontrollawnew}) in (\ref{eqfinalform}), the uncertainty output can be written as follows:
\begin{equation}
\begin{split}
\tilde{z}&=\tilde{K}x+\tilde{G}(G x+v), \\
&=(\tilde{K}+\tilde{G}G) x+ \tilde{G}v.
\end{split}
\end{equation}

Finally, we can write the closed loop system considering saturation uncertainty with antiwindup signal $v$ as follows:
\begin{equation}
\label{eqfinalformlqg}
\begin{split}
\dot{x}&= \bar{A}x(t)+ B_1 v+ \tilde{B}_2\zeta,\\
\tilde{z}&=\tilde{C}_1 x + \tilde{D}_1 v,\\
\tilde{y}&= \tilde{C}_2 x+ \tilde{D}_2 \zeta,
\end{split}
\end{equation}
where, $\bar{A}=(A-\bar{B}G)$, $\tilde{C}_1=\tilde{K}+\tilde{G}G$, $\tilde{D}_2=[\mathbf{0}\quad \mathbf{I}]$, $B_1=\bar{B}$, $\tilde{C}_2=\mathbf{0}$,
\[
\tilde{B}_2=\left[\begin{array}{ccccc}
C_1 & C_2 & \cdots & C_l & -B
\end{array}\right],
\]
\[
\zeta=\left[\begin{array}{ccccc}
\zeta_1 & \zeta_2 & \cdots & \zeta_l & \hat{w}(\cdot)
\end{array}\right]^T.
\]

\subsection{Minimax LQG controller synthesis for antiwindup augmentation}
We now design an antiwindup controller of the form (\ref{eqawcontrol}) using minimax LQG design procedure. 
The design procedure for the  standard minimax LQG is given in \cite{IP}. Here, we present a summary of the method and then present our approach to designing a minimax LQG antwindup controller.
The controller (\ref{eqawcontrol}) can be designed after stabilizing the system (\ref{eqgform1}) using (\ref{eqcontrollaw}) and writing the closed loop system in the form (\ref{eqfinalformlqg}). The minimax LQG control problem \cite{IP} involves  finding a controller which minimizes the maximum value of the following cost function:
\begin{equation}
\label{eqfcost}
J=\lim_{T\rightarrow\infty} \frac{1}{2T}\mathbf{E}\int_0^{T}(x(t)^T Q x(t)+v(t)^T R v(t))dt,
\end{equation}
where $R>0$ and $Q>0$.
The maximum value of the cost is taken over all uncertainties satisfying the IQC (\ref{eqIQC}).
If we define a variable
\begin{equation}
\label{eqPsi}
{\psi}=\left[\begin{array}{c}
Q^{1/2} x\\
R^{1/2} v
\end{array}\right],
\end{equation}
the cost function (\ref{eqfcost}) can be written as follows:
\begin{equation}
\label{eqffcost}
J=\lim_{T\rightarrow\infty} (\frac{1}{2T})\mathbf{E}\int_0^{T}\Vert \psi \Vert^2 dt.
\end{equation}
\begin{figure}[t]
\hfill
\begin{center}
\epsfig{file=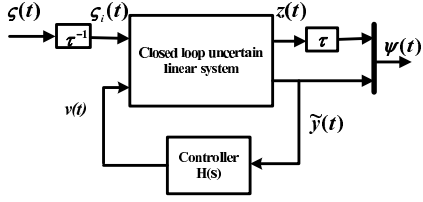, scale=0.5}
\caption{Scaled $H_\infty$ control problem.}
\label{fig:scaled_risk}
\end{center}
\end{figure}

The minimax optimal controller problem can now be solved by solving a scaled risk-sensitive control problem \cite{IP} which corresponds to a scaled $H_\infty$ control problem; e.g. see \cite{IP_LQG}. The scaled risk-sensitive control problem considered here (see Fig. \ref{fig:scaled_risk}) allows for a tractable solution in terms of the following pair of $H_{\infty}$ type algebraic Riccati equations for ${C}_2=0$.:
\begin{align}
\label{eqARE1}
\bar{A}Y_{\infty}&+Y_{\infty}\bar{A}^T + Y_{\infty}(\tau^{-1}R_{\tau})Y_{\infty} \nonumber\\
&+ \tilde{B_2}(I-\tilde{D_2}^T  \Gamma^{-1} \tilde{D_2})\tilde{B_2}^T =0,
\end{align}
and
\begin{align}
\label{eqARE2}
&X_{\infty}(\bar{A}-B_1 G_{\tau}^{-1}  \Upsilon_{\tau}^T )+(\bar{A}-B_1 G_{\tau}^{-1} \Upsilon_{\tau}^T )^T  X_{\infty}\nonumber \\
&- X_{\infty}(B_1 G_{\tau}^{-1}B_1^T -\tau^{-1}\tilde{B_2} \tilde{B_2}^T )X_{\infty} \nonumber \\
&+ (R_{\tau}-\Upsilon_{\tau} G_{\tau}^{-1}\Upsilon_{\tau}^T )=0,
\end{align}
where,
\[
R_{\tau}\triangleq Q+\tau \tilde{C_1}^T  \tilde{C_1},\quad G_{\tau} \triangleq R+\tau \tilde{D_1}^T  \tilde{D_1},\quad \Gamma_{\tau}\triangleq \tau \tilde{C_1}^T  \tilde{D_1}.
\]
In order to obtain solutions to both of the algebraic Riccati equations $Y_{\infty} > 0$, $X_{\infty}>0$, the system (\ref{eqfinalformlqg}) is required to satisify the following assumption: 
\begin{assumption}{\ \\}
\label{assump}
\begin{itemize}
\item [1.] The matrix $\tilde{C}_1$ and $\tilde{D}_1$ satisfy the condition $\tilde{C}_1^T\tilde{D}_1=0$.
\item [2.] The matrix $\tilde{D}_2$  satisfies the condition $\tilde{D}_2^T\tilde{D}_2>0$.
\item [3.] $\bar{A}$ is Hurwitz.
\item [4.] The pair $(\bar{A},\tilde{B}_2)$ is stabilizable and $\tilde{B}_2\neq0$
\item [5.]    The pair $(\bar{A},\tilde{B}_1)$ is stabilizable and $y(t)$ is measurable.
\item [6.]    The matrix $\tilde{B}_2$ and $\tilde{D}_2$ satisfy the condition $\tilde{B}_2\tilde{D}_2^T=0$.
\item [7.]     $R_{\tau}-\Upsilon_{\tau} G_{\tau}^{-1} \Upsilon_{\tau}^T \geq 0$
\item [8.]    The pair ($\bar{A}-\tilde{B}_1 G_{\tau}^{-1} \Upsilon_{\tau}^T , R_{\tau}-\Upsilon_{\tau} G_{\tau}^{-1} \Upsilon_{\tau}^T $) is detectable.
\item [9.]     The pair ($\bar{A}, \tilde{B}_2(\mathbf{I}-\tilde{D}_2^T \Gamma^{-1} \tilde{D}_2)$) is stabilizable.
\end{itemize}
\end{assumption}
If the solutions of the Ricatti equations satisfy $I-\tau^{-1} Y_{\infty} X_\infty > 0$ and the parameter $\tau > 0$ is chosen such that it minimizes the cost bound ($W_\tau$) defined by
\begin{align}
\label{eqbound}
&W_\tau \triangleq tr[(\tilde{B_2} \tilde{D_2}^T)(D_2D_2^T)^{-1}\nonumber\\
&\times (\tilde{D_2} \tilde{B_2}^T)X(I-YX)^{-1}+\tau Y R_{\tau}],
\end{align}
then the antiwindup controller  matrices in (\ref{eqawcontrol}) can be obtained as follows:
\begin{align*}
C_{aw}&=-G_{\tau}^{-1}(\tilde{B}_1^T  X_{\infty}+\Upsilon_{\tau}^T );\\
B_{aw}&=(I-\tau^{-1} Y_{\infty} X_\infty )^{-1}(\tilde{B}_2 \tilde{D}_2^T )\Gamma^{-1};\\
A_{aw}&=\tilde{\bar{A}}+\tilde{B}_1 C_{aw} + \tau^{-1}(\tilde{B}_2-B_c \tilde{D}_2)\tilde{B}_2^T  X_{\infty}.
\end{align*}

\begin{theorem}
\label{th1}
Consider the uncertain linear system (\ref{eqfinalformlqg}) with the cost function (\ref{eqfcost}) and suppose assumption \ref{assump} is satisfied.  Then the controller (\ref{eqawcontrol}) minimzes the bound on the cost function (\ref{eqfcost}) such that $W_\tau(\cdot)= \inf_{v(\cdot)} \sup J(\cdot)$ and guarantees the stability of the system (within certain domain of attraction) in the presence of saturation nonlinearity if the following conditions are hold for an arbitrary $\tau>0$:
\begin{itemize}
\item [1.] The algebraic Ricatti equation (\ref{eqARE1}) admits a minimal positive-definite solution $Y_{\infty}$. 
\item [2.] The algebraic Ricatti equation (\ref{eqARE2}) admits a minimal nonnegative-definite solution $X_{\infty}$.
\item [3.] The matrix $\mathbf{I}-\begin{small}\frac{1}{\tau}\end{small} Y_{\infty}X_{\infty}$ has only positive eigenvalues; that is the spectral radius of the matrix $Y_{\infty}X_{\infty}$ satisfies the condition $\rho(Y_{\infty}X_{\infty})<\tau$.
\end{itemize}
\end{theorem}
\startproof 
See \cite{IP}.\finishproof

\section{Example}\label{sec:example}
In this section, we apply our proposed antiwindup synthesis approach to design a velocity and attitude tracking controller with antiwindup augmentation for an air-breathing hypersonic flight vehicle (AHFV). This design example has been taken from our previous work \cite{Rehman_GNC01, Rehman_MSC01} and it is observed that the AHFV system is subject to actuator saturation. Here, we use the uncertain linearized model of AHFV which was obtained using the robust feedback linearization method in \cite{Rehman_MSC01}. The linearized model is 2-input and 2-output system which is subject to $24$ uncertainty parameters $p_1, ~p_2,~\cdots,~p_{24}$. For the ease of reference the corresponding linearized model of the form (\ref{eqgform1}) is shown below:

\begin{align}
\label{equmodel}
\dot {\chi}(t) &=A \chi(t)+B \bar{v}(t)+\sum^2_{j=1} C_j\zeta_ j;\\
z_i(t)&=K_i{\chi(t)}+G_i \bar{v}(t);\quad i=1,2\\
y(t)&=\bar{C}_2 \chi(t)+ D_2 \bar{v}(t)
\end{align}
where,
\[
A=\left[\begin{array}{ccccccccc}
0 & 1 & 0 & 0 & 0 & 0 & 0 & 0 & 0\\
0 & 0 & 1 & 0 & 0 & 0 & 0 & 0 & 0\\
0 & 0 & 0 & 1 & 0 & 0 & 0 & 0 & 0\\
0 & 0 & 0 & 0 & 0 & 0 & 0 & 0 & 0\\
0 & 0 & 0 & 0 & 0 & 1 & 0 & 0 & 0\\
0 & 0 & 0 & 0 & 0 & 0 & 1 & 0 & 0\\
0 & 0 & 0 & 0 & 0 & 0 & 0 & 1 & 0\\
0 & 0 & 0 & 0 & 0 & 0 & 0 & 0 & 1\\
0 & 0 & 0 & 0 & 0 & 0 & 0 & 0 & 0\\
\end{array}\right],~~
B=\left[\begin{array}{cc}
0 & 0  \\
0 & 0  \\
0& 0  \\
1 & 0  \\
0 & 0  \\
0 & 0  \\
0 & 0 \\
0 & 0  \\
0 & 1 \\
\end{array}\right];
\]
\[
C_1=\left[\begin{array}{ccccc}
0 & 0 &\cdots & 0 & 0\\
0 & 0 &\cdots & 0 & 0 \\
0 & 0 &\cdots & 0 & 0  \\
1 & 1 &\cdots & 1 & 1\\
0 & 0 &\cdots & 0 & 0\\
\vdots & \vdots & \ddots & \vdots &\vdots\\
0 & 0 &\cdots & 0 & 0
\end{array}\right]
,\quad
C_2=\left[\begin{array}{ccccc}
0 & 0 &\cdots & 0 & 0\\
0 & 0 &\cdots & 0 & 0 \\
0 & 0 &\cdots& 0 & 0  \\
0 & 0 &\cdots & 0 & 0  \\
\vdots & \vdots & \ddots & \vdots &\vdots\\
0 & 0 &\cdots & 0 & 0\\
1 & 1 &\cdots & 1 & 1
\end{array}\right];
\]
\[
K_1=\left[\begin{array}{lc}
\Delta \tilde{w}_{1(p_1)} \\
\Delta \tilde{w}_{1(p_2)} \\
\vdots \\
\Delta \tilde{w}_{1(p_{23})}\\
\Delta \tilde{w}_{1(p_{24})} \\
\end{array}\right],\quad
K_2=\left[\begin{array}{lc}
\Delta \tilde{w}_{2(p_1)} \\
\Delta \tilde{w}_{2(p_2)} \\
\vdots \\
\Delta \tilde{w}_{2(p_{23})}\\
\Delta \tilde{w}_{2(p_{24})} \\
\end{array}\right];
\]
\[
G_1=\left[\begin{array}{lc}
\Delta \check{w}_{1(p_1)} \\
\Delta \check{w}_{1(p_2)} \\
\vdots \\
\Delta \check{w}_{1(p_{23})}\\
\Delta \check{w}_{1(p_{24})} \\
\end{array}\right],\quad
G_2=\left[\begin{array}{lc}
\Delta \check{w}_{2(p_1)} \\
\Delta \check{w}_{2(p_2)} \\
\vdots \\
\Delta \check{w}_{2(p_{23})}\\
\Delta \check{w}_{2(p_{24})} \\
\end{array}\right],
\]
and  $\chi(t)\in \mathbb{R}^9$ is the state vector, and $\bar{v}(t)=[\bar{v}_1~\bar{v}_2]^T\in\mathbb{R}^2$ is the control input vector.
Note that the size of matrix $Cj$ is $9 \times 24$. Also, $\zeta_ 1\in \mathbb{R}^{24}$, and $\zeta_2\in \mathbb{R}^{24}$ are uncertainty inputs, $z_1(t)\in \mathbb{R}^{24}$, and $z_2(t)\in \mathbb{R}^{24}$ are the uncertainty outputs, $\Delta \check{w}_{(\cdot)}$ and $\Delta \tilde{w}_{(\cdot)}$ represent the magnitude of the uncertainties in the system.

\begin{figure}[t]
\hfill
\begin{center}
\epsfig{file=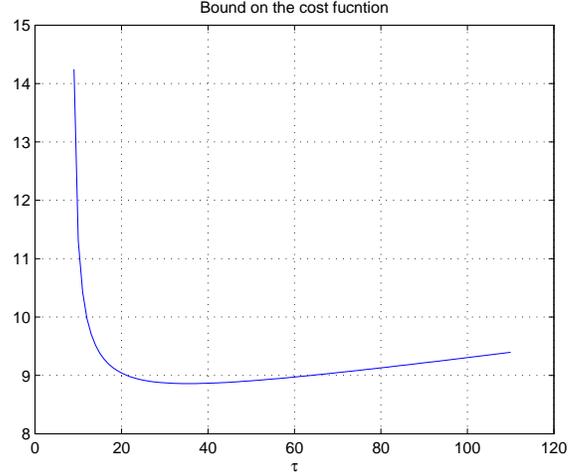, scale=0.6}
\caption{Bound on the cost function with varying $\tau$.}
\label{fig:costbound}
\end{center}
\end{figure}

As a first step, a controller of the from (\ref{eqcontrollaw}) is designed which gives a stable close loop system in the absence of actuator saturations as discussed in the Section \ref{sec:system}. In the second step, we augment the controller with antiwindup compensation (\ref{eqcontrollawnew}) so that the performance degradation is minimized and system remains stable in the selected domain of attraction as discussed in Section \ref{sec:awproblem}. The antiwindup compensator of the form (\ref{eqawcontrol}) is obtained by selecting appropriate state and control weighting matrices $Q$ and $R$.
For this example, the following parameters are selected to obtain a good antiwindup controller:
\begin{equation}
\label{eqSW}
\mathbf{Q}=1000 \times \textbf{diag}\left[
\begin{array}{c}
1, 1, 1, 1,1,1, 1, 1, 1
\end{array}\right],
\end{equation}
\begin{equation}
\label{eqCW}
\mathbf{R}=\left[
\begin{array}{cc}
3 & 0\\
0 & 8
\end{array}\right],\quad \tau=35.
\end{equation}
The parameter $\tau$ has been selected which gives the minimum cost bound (\ref{eqbound}) as shown in Fig \ref{fig:costbound}. 
\begin{remark}
The selection of state and control weighting matrices plays a significant role in the synthesis of the minimax LQG antiwindup controller. The selection should be made such that the dynamic gain of the closed loop transfer function should not be too high or too low. The antiwindup signal $v$ should be smaller than the actual input $u$.
\end{remark}
\subsection{Simulation Results}
\begin{figure}[t]
\hfill
\begin{center}
\epsfig{file=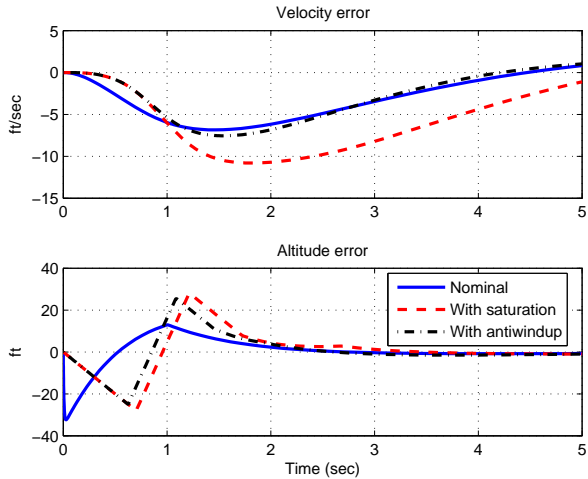, scale=0.6}
\caption{Velocity and altitude tracking error responses. The solid line shows the response with the nominal model, `- -' shows the response with input saturation, `-.' shows the response with antiwindup augmentation.}
\label{fig:result11}
\end{center}
\end{figure}

\begin{figure}[t]
\hfill
\begin{center}
\epsfig{file=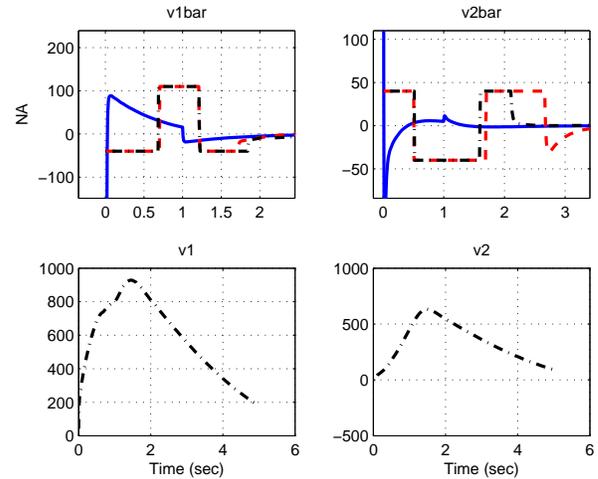, scale=0.6}
\caption{Controller output responses. The solid line shows the response with the nominal model, `- -' shows the response with input saturation, `-.' shows the response with antiwindup augmentation.}
\label{fig:result12}
\end{center}
\end{figure}

Simulation results using the antiwindup augmentation procedure discussed above are shown in Fig. \ref{fig:result11}- Fig. \ref{fig:result12}. The solid (blue) line shows the response of the nominal system without actuator saturation, the dashed (red) line shows the response with the actuator saturation and the dashed-dot (black) line shows the response with the antiwindup augmentation compensator. The results show that in the presence of actuator saturation, the antiwindup augmentation removes the actuator saturation degradation in a very effective way. The tracking errors remain small and bounded for both the cases of velocity and altitude reference input commands.
\section{Conclusion}\label{sec:concl}
In this paper, a minimax linear quadratic Gaussian (LQG) antiwindup augmentation compensator has been proposed for an uncertain linear plant subject to input saturation. The design employs a two-stage process in which a robust controller is designed using minimax linear quadratic regulator (LQR) without considering actuator saturation as the first step. Then antiwindup augmentation is provided in the second step. The proposed approach has been applied to a tracking control problem for an air-breathing hypersonic flight vehicle (AHFV) system. Simulation results show that the proposed approach is very effective in dealing with actuator saturation. It is observed that the proposed antiwindup augmentation,  reduces the degradation in performance. Antiwindup designs for  nonlinear uncertain systems using feedback linearization are areas for future research.

\section{ACKNOWLEDGMENTS}
This research was supported by the Australian Research Councils and Australian Space Research Program.




\end{document}